\documentstyle[preprint,epsfig,aps]{revtex}

\draft
\begin{document}
%\wideabs{
\title{A study of Peierls instabilities for a two-dimensional $t$-$t'$ model}
\author{Qingshan Yuan$^{1,2}$}
\address{$^1$ Experimentalphysik VI, Universit\"at Augsburg, 
 86135 Augsburg, Germany\\
$^2$Pohl Institute of Solid State Physics, Tongji University, 
Shanghai 200092, P.R.China}

\maketitle

\begin{abstract}
In this paper we study Peierls instabilities for 
a half-filled two-dimensional tight-binding model with 
nearest-neighbour hopping $t$ and next nearest-neighbour hopping $t'$
at zero and finite temperatures. Two dimerization patterns corresponding to 
the same phonon vector $(\pi, \pi)$ are considered to be realizations of 
Peierls states. The effect of imperfect nesting
introduced by $t'$ on the Peierls instability, the properties of the
dimerized ground state, as well as the competition between two 
dimerized states for each $t'$ and temperature $T$, are investigated. 
It is found: (i). 
The Peierls instability will be frustrated by $t'$ for each of the
dimerized states. The Peierls transition itself,
as well as its suppression by $t'$, may be of second- or first-order. 
(ii). When the two dimerized states are considered jointly, 
one of them will dominate the other depending on parameters $t'$
and $T$. Two successive Peierls transitions, that is, the system passing
from the uniform state to one dimerized state and then to the other 
take place with decrease of temperature for some $t'$ values.  
Implications of our results to real materials are discussed.
\end{abstract}


%}

\section{Introduction}

The Peierls instability, associated with bond-order or 
charge-density waves, is an important phenomenon in low dimensional materials
driven by electron-phonon interaction \cite{Gruner}.
The occurance of such an instability
is directly relevant to the nesting property of the Fermi surface
(FS) in the normal high temperature state of materials. 
It has been observed in quasi-one 
dimensional (1D) metals such as polyacetylene (CH)$_x$,
blue bronzes A$_{0.3}$MoO$_3$ (A=K, Rb, Tl) \cite{Schlenk}, 
as well as quasi-two dimensional (2D) materials such as 
purple bronzes AMo$_6$O$_{17}$ (A=Na, K, Tl) \cite{Schlenk,Dumas,Whangbo,Qin}
and a series of monophosphate tungsten bronzes (PO$_2$)$_4$(WO$_3$)$_{2m}$ 
with $4\le m\le 14$ \cite{Schlenker,Wang,Hess,Beille}.

\begin{figure}[ht]
\epsfxsize=12cm
\epsfysize=6cm
\centerline{\epsffile{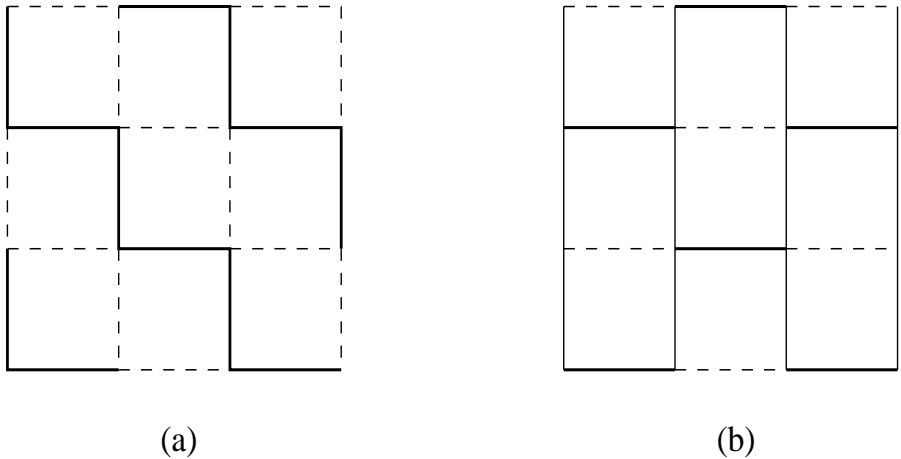}}
\medskip
\caption{The lattice distortion patterns (a) and (b). In the figure 
a thick solid line 
corresponds to a strong bond with hopping integral $t(1+\delta)$, a dashed 
line corresponds to a weak bond with hopping integral $t(1-\delta)$, and a 
thin solid line corresponds to a normal bond with hopping integral $t$. Both 
patterns correspond to phonons with wave vector $(\pi,\pi)$. The dimerization 
is along two axes for pattern (a), while only along the 
$x$ axis for pattern (b).}
\label{Fig:Pattern}
\end{figure}

Theoretically the description for the Peierls instability was mainly developed
for 1D metals, and relatively few investigations are
concerned with 2D systems \cite{Tang,Mazumdar,Clay,Ono,Lin,Yuan}. 
Actually the 2D study of the Peierls instability 
is not straightforward because of the more complex FS structure
as well as richer distortion patterns. For example, in Fig. \ref{Fig:Pattern}
we have shown two possible dimerization patterns for the lattice distortion 
and the corresponding bond hopping in two dimensions.
Both of them correspond to phonons with wave vector $(\pi,\pi)$, 
and their difference is that for pattern (a) the dimerization is in both
directions, while it is only in one direction for pattern (b).
The simplest model to discuss the Peierls instability in two dimensions is
a square lattice tight-binding model with nearest-neighbour (n.n.) 
hopping $t$, i.e., the 2D version of the 
well-known Su-Schrieffer-Heeger (SSH) model \cite{SSH}. It was originally
studied by several authors in connection to high-$T_c$ 
superconductors \cite{Tang,Mazumdar}. At half-filling
the FS is perfectly nested with nesting vector $Q=(\pi,\pi)$ so that 
arbitrarily small electron-lattice 
coupling \cite{Yuan} will induce a Peierls instability into
any one of the dimerized states shown in Fig. \ref{Fig:Pattern} 
(if quantum lattice fluctuations are ignored).

The above situation, i.e., perfect nesting of FS is, however, special 
because it may be easily
broken for example, by introduction of next nearest-neighbour (n.n.n.) 
hopping $t'$. Actually for real materials, e.g., quasi-2D high-$T_c$ 
superconductors n.n.n. hopping was found to be important. 
Also for those quasi-2D materials which show a Peierls instability, perfect
nesting of their Fermi surfaces is not satisfied. It is reasonable to 
adopt a tight-binding model with n.n. and n.n.n. hopping to simulate an 
essential property of them: the relevance of nesting. Then a question is 
naturally asked: how is the Peierls instability affected by $t'$?

It is the purpose of this paper to make a systematical study of 
Peierls instabilities for a 2D
$t$-$t'$ model, where the nesting property of the FS can be modulated.
In contrast, the corresponding problem in the 1D case might be less attractive
because the FS is always composed of two points and perfectly nested as long
as $t'/t<0.5$, and then at least in this region $t'$ has no effect on 
Peierls instability \cite{Qing}. The 2D problem at zero temperature
was addressed in an earlier work \cite{Yuan}. It was found that 
the Peierls instability will be suppressed with increasing $t'$
for each of the two patterns (a) and (b), when they are studied individually. 
Moreover, when they are considered jointly, one of them
was found to dominate the other or reverse when $t'$ is less
or greater than some critical value $t'_{ab}$, i.e.,
there exists a crossover between the two patterns at $t'=t'_{ab}$. 
Further one may conjecture that the
same behavior should also occur at finite temperatures, and most probably the
value $t'_{ab}$ will change with temperature. Then an intriguing phenomenon
may arise: at some $t'$, the two dimerized states will
dominate successively with change of temperature so that double Peierls 
transitions (one is between uniform state and dimerized state, the other
is between two dimerized states) are expected to take place as a function of 
temperature. Actually, two successive Peierls transitions 
were observed in TlMo$_6$O$_{17}$ \cite{Qin} and 
most of the series (PO$_2$)$_4$(WO$_3$)$_{2m}$
\cite{Schlenker,Wang,Hess,Beille}. 
So it is interesting to extend the previous work to 
finite temperatures to discover the above phenomenon.
At the same time, new findings at ground state will be supplemented.

Before entering the details we implement a remark on the lattice distortion. 
Recently, Ono and Hamano studied the 2D Peierls 
instability numerically in the absence of $t'$ and found that a complex 
multi-mode distortion pattern has a lower ground state energy than any one of 
the $(\pi,\pi)$ structure presented here \cite{Ono}.
As also described by the authors themselves, however, such a distortion pattern
is not unique and may be infinitely degenerate. So it is difficult to extract
an explicit pattern for real applications. Moreover, no further confirmation
on these new patterns is present at the moment. Thus the question of 
the real lattice distortion for the 2D Peierls system is still open 
even in the case of $t'=0$. Under these circumstances we prefer to adopt the 
two commonly assumed dimerization patterns (a) and (b) 
as candidates of the Peierls state. These two patterns comply with the nesting
vector $Q=(\pi,\pi)$ and realize an unconditional Peierls instability 
at $t'=0$. Similarly, one may also argue that for each $t'$ and temperature 
there may exist a lattice distortion structure, 
though perhaps extremely complicated, with the lowest (free) energy.
We have no intention to include all kinds of lattice distortions, 
which is actually impossible in an analytical treatment on an infinite 
lattice. Instead, throughout the work we constrain the lattice distortions to 
the patterns (a) and (b) (although a discussion on a generalization to these
two patterns is given in the last section), 
and based on this, we intend to investigate the 
presence of $t'$ after fully understanding the $t'=0$ case. 
This is a natural topic by physical consideration as illustrated in the 
preceding paragraphs, which is helpful to understand the 2D 
Peierls instability itself as well as relevant materials 
even if the assumed dimerized 
states (a) and (b) might be not the lowest energy Peierls states. 
%The significance is also shown by the rich outcome discovered below.

\section{Formulation}

Our Hamiltonian represents a half filled 2D $t-t'$ model including
lattice displacements. It reads \cite{Yuan}:
\begin{eqnarray}
H & = & -t\sum_{i,j,\sigma} [1+\alpha (u_{i,j}^x-u_{i+1,j}^x)]
  (c_{i,j,\sigma}^{\dagger}c_{i+1,j,\sigma}+{\rm h.c.})
-t\sum_{i,j,\sigma} [1+\alpha (u_{i,j}^y-u_{i,j+1}^y)]
  (c_{i,j,\sigma}^{\dagger}c_{i,j+1,\sigma}+{\rm h.c.}) \nonumber\\
& & -t'\sum_{i,j,\sigma}(c_{i,j,\sigma}^{\dagger}c_{i+1,j+1,\sigma}+
  c_{i,j,\sigma}^{\dagger}c_{i+1,j-1,\sigma}+ {\rm h.c.})+
{K\over 2} \sum_{i,j}[(u_{i,j}^x-u_{i+1,j}^x)^2+(u_{i,j}^y-u_{i,j+1}^y)^2]\ .
\label{eq:H}
\end{eqnarray}
All above notations are conventional: $c_{i,j,\sigma}^{\dagger} 
(c_{i,j,\sigma})$ is the creation 
(annihilation) operator for an electron at site $(i,j)$ with spin $\sigma$
($i$ denotes $x$ coordinate and $j$ denotes $y$ coordinate);
$u_{i,j}^{x/y}$ is the displacement component of site $(i,j)$ 
in $x/y$ direction; $t,\ t'$ are n.n and n.n.n. hopping parameters 
and $\alpha$ is the electron-lattice coupling constant. The last term above 
describes the lattice elastic potential energy with $K$ the elastic constant.
The phonons are treated in adiabatic approximation. 

The lattice distortion, as shown in Fig. \ref{Fig:Pattern}, may be 
explicitly expressed as  
$$u_{i,j}^x-u_{i+1,j}^x=(-1)^{i+j}u,\ \
  u_{i,j}^y-u_{i,j+1}^y=(-1)^{i+j}u
$$
for pattern (a) and
$$u_{i,j}^x-u_{i+1,j}^x=(-1)^{i+j}u,\ \  
  u_{i,j}^y-u_{i,j+1}^y=0\ \ \ \ \ 
$$
for pattern (b), where $u$ is the amplitude of dimerization which needs to
be determined. Experimentally these two patterns may be differentiated
by high resolution x-ray or neutron scattering.
For convenience, two dimensionless
parameters are defined as follows: the dimerization 
amplitude $\delta=\alpha u$ and the electron-lattice coupling 
constant $\eta= \alpha^2 t/K$.

The Hamiltonian (\ref{eq:H}) may be easily diagonalized to give the 
following electronic spectra in momentum space for patterns (a) and
(b), respectively,
\begin{eqnarray}
\varepsilon_{{\bf k},a}^{\pm} & = & -4t'\cos k_x \cos k_y
\pm 2\sqrt{(\cos k_x+\cos k_y)^2+\delta^2(\sin k_x+\sin k_y)^2}\ , \nonumber\\
\varepsilon_{{\bf k},b}^{\pm} & = & -4t'\cos k_x \cos k_y
 \pm 2\sqrt{(\cos k_x+\cos k_y)^2+\delta^2\sin^2 k_x}\ .
\label{eq:spm}
\end{eqnarray}

Before continuing we check the symmetries of the above spectra. 
It is easy to see that the spectra in both cases are
invariant under the same operation: $k_x\rightarrow k_x\pm \pi$ and
simultaneously $k_y\rightarrow0 k_y\pm \pi$. In addition, the spectra 
$\varepsilon_{{\bf k},a}^{\pm}$ have reflection symmetry:
$k_x\rightarrow -k_x$ and simultaneously $k_y\rightarrow -k_y$, as well as 
exchange symmetry: $k_x\leftrightarrow k_y$; while the 
$\varepsilon_{{\bf k},b}^{\pm}$ have only the reflection symmetry:
$k_x\rightarrow -k_x$ {\it or} $k_y\rightarrow -k_y$, or both.

The electronic grand canonical partition function $\Xi$ may be written as:
\begin{equation}
\ln \Xi =2\sum_{{\bf k},\nu} \ln [1+e^{-\beta 
(\varepsilon_{{\bf k}}^{\nu}-\mu)} ]
\end{equation}
with $\beta =1/k_B T$ ($T$: temperature) and $\nu=\pm$: band index.
The factor $2$ in front of the summation includes spin degeneracy. 
The chemical potential $\mu$ is adjusted to yield
the right filling, i.e., $N_e={1\over \beta}
{\partial \over \partial \mu}\ln \Xi$. $N_e$ is the total electron number,
equal to the total number of lattice sites $N$ at half-filling.

The required optimal dimerization parameter $\delta^*$ is determined by
minimization of the total free energy which is given by 
\begin{equation} 
F=-{1\over \beta} \ln \Xi + N_e \mu + E_L\ .
\end{equation}
Here $E_L$ denotes the lattice elastic energy which is, in unit of $t$,
$N\delta^2/\eta$ for
pattern (a) and $N\delta^2/(2\eta)$ for pattern (b). Note that $E_L$
for (a) is twice of that for (b) under the same $\delta$.

Throughout the paper we take the Bolzmann constant $k_B=1$ and the n.n. 
hopping integral $t$ as the energy unit. The parameter $\eta$ is fixed at 
a typical value $0.5$. The variation of $\eta$ should not change the results
qualitatively, as has been seen at ground state in the earlier work 
\cite{Yuan}. Also notice that
the Hamiltonian with $-t'$ may be mapped onto that with $t'$ at half-filling
through the transformation: $c_{i,j,\sigma}\rightarrow 
(-1)^{i+j}c_{i,j,\sigma}^{\dagger}$, 
so we only consider $t'>0$ in the following.

\section{Results at $T=0$}

The effects of $t'$ on the Peierls instability were carefully 
studied at ground state in Ref. \cite{Yuan}, and
the main results are summarized here. For each of the two patterns, 
it was found that $t'$ tends to suppress the Peierls instability. 
But the details in both cases are 
different: for (a) the suppression is of first-order 
while for (b) it is of second-order. The critical values of $t'$
for suppression in both cases are close but slightly different.
More interestingly, a crossover 
between the two dimerized states themselves may take place
with change of $t'$.

\begin{figure}
\epsfxsize=14cm
\epsfysize=5.8cm
\centerline{\epsffile{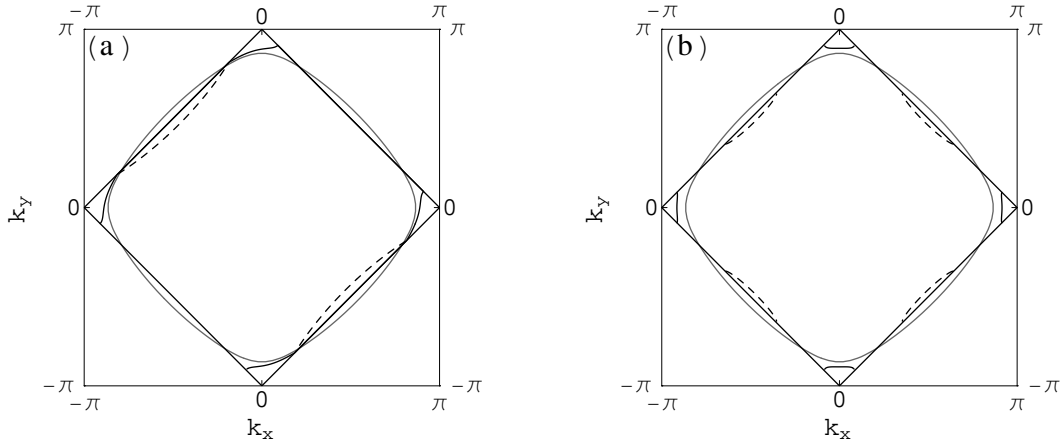}}
\medskip
\caption{The Fermi surfaces before dimerization (gray curve) and after 
dimerization (solid and dashed curves) for $t'=0.15$
for both patterns. For pattern (a), the FS of the `$+$' 
band (dashed curve) is
formed just by the two pockets around $(\pi/2,-\pi/2)$ and $(-\pi/2,\pi/2)$;
while for pattern (b) it is formed by the four pockets. The `$-$' band for 
each case has the large FS (solid curve) with nearly the square shape, 
except for the cut corners. The square: $|k_x|+|k_y|=\pi$ is
the reduced Brillouin zone (also the FS in the case of $t'=\delta =0$).}
\label{Fig:FS}
\end{figure}

After a brief presentation of the above results, our main attention in this 
section is paid to other useful properties in the ground state, which were  
not discovered previously. First let us have a look at the variation 
of the FS before and after dimerization. This helps thoroughly 
understand the impact of the FS structure on the Peierls instability.
We take $t'=0.15$ for example and plot the Fermi surfaces in Fig. \ref{Fig:FS}
for both patterns (also cf. the plots of electronic spectra shown 
later). In this figure, the FS before 
dimerization is shown by the gray curve in the original 
Brillouin zone, while after dimerization the Fermi surfaces for the `$-$'
and `$+$' bands are shown in the reduced Brillouin zone by the solid
and dashed curves, respectively. Please note by the way the differences of 
the FS structures for both patterns which are controlled by the symmetries of 
the respective spectra addressed above. From Fig. \ref{Fig:FS} a common 
observation in both patterns is that after dimerization there are a few 
electrons resting in the `$+$' band, whose energy levels are lifted as 
seen from Eq. (\ref{eq:spm}). That is, the energies for this small fraction 
of electrons are increased by dimerization, leading to a partial cancellation 
of the energy gain 
acquired by most of electrons dropping into the `$-$' band. This situation is 
in contrast to that for $t'=0$
where the original FS is perfectly nested so that all electrons drop into
the `$-$' band with each energy level reduced (at least not raised) 
by dimerization. Therefore, one may see that
the imperfect nesting of the original FS induced by $t'$ is unfavorable to 
gain electronic energy {\it to the largest extend}. And 
once such a gain is not sufficient to compensate the cost of
the elastic energy, the Peierls instability will be prohibited.

\begin{figure}
\epsfxsize=12cm
\epsfysize=10.5cm
\centerline{\epsffile{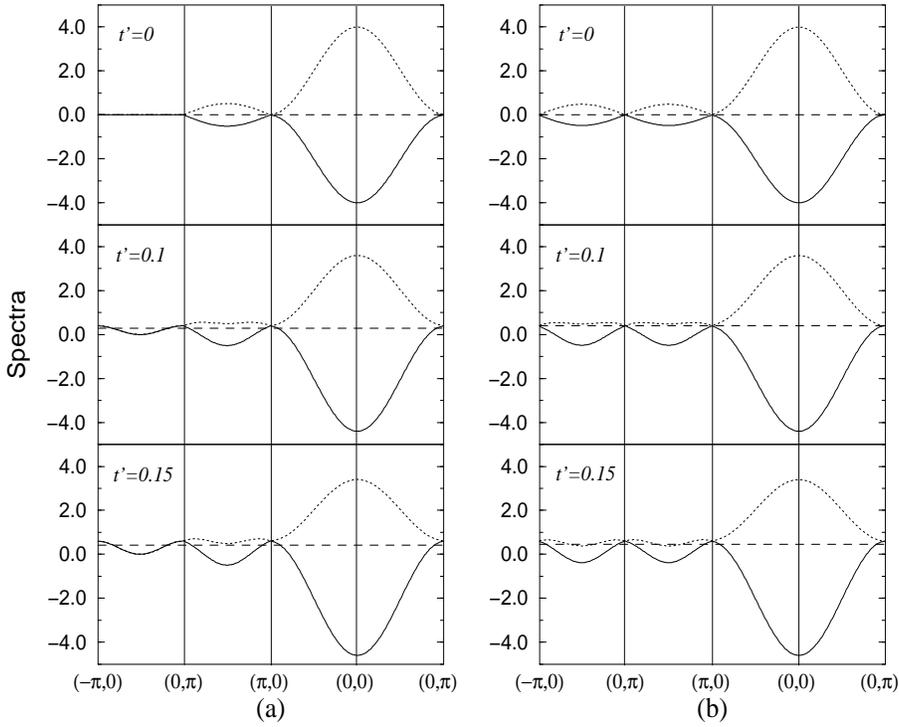}}
\medskip
\caption{The electronic spectra in the dimerized states for several $t'$
values for both patterns. The solid and dotted curves in each panel represent
energy bands `$-$' and `$+$', respectively, and the dashed horizontal line 
shows the chemical potential. Note that the two bands in the 
region $(-\pi,0)$-$(0,\pi)$ are identical for pattern (a). 
%The related parameters are listed: i. For
%pattern (a), $t'=0/0.1/0.15$, $\delta^*=0.1281/0.1255/0.1229$, 
%$\mu=0/0.282/0.4165$. ii. For pattern (b), $t'=0/0.1/0.15$, 
%$\delta^*=0.2422/0.2422/0.19$, $\mu=0/0.4/0.4556$.
}
\label{Fig:SPM}
\end{figure}

\begin{figure}
\epsfxsize=12cm
\epsfysize=9cm
\centerline{\epsffile{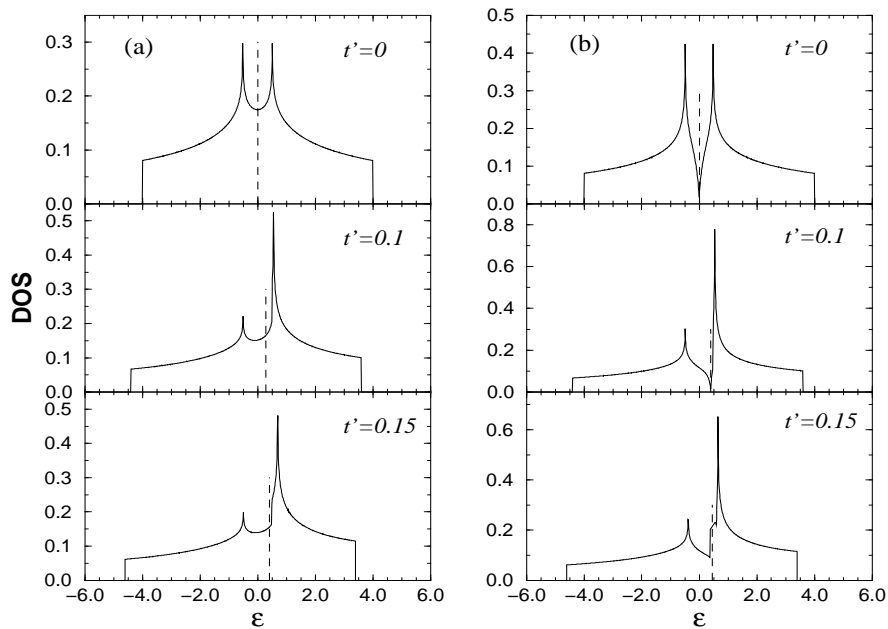}}
\medskip
\caption{The density of states (DOS) for each case corresponding to Fig.
\ref{Fig:SPM}. The dashed vertical line in each panel shows the chemical
potential.}
\label{Fig:DOS}
\end{figure}

Another observation we want to illustrate is that the dimerized state 
(a) or (b), due to the Peierls transition, is metallic or semi-metallic
in two dimensions even for $t'=0$ rather than insulating as in one dimension. 
Actually there is no finite gap opened by the dimerization in any case. 
To more clearly see this point 
we have plotted the electronic spectra in the dimerized states
for several $t'$ values in Fig. \ref{Fig:SPM}, as well as 
their corresponding density of states 
(DOS), i.e., ${1\over N}\sum_{{\bf k},\nu}\delta (\varepsilon -
\varepsilon_{{\bf k}}^{\nu})$
in Fig. \ref{Fig:DOS}. 
The two bands are found to touch or overlap for each $t'$ and 
for each pattern, as also seen from DOS where no gap appears in each case. 
We point out that the Peierls instability, when occuring in those 
real quasi-2D materials, is always associated with a {\it metal-metal} 
transition \cite{Schlenker}.

At this stage, it is worthwhile to mention several 
points seen from Fig. \ref{Fig:SPM}: i. The two bands in the 
region $(-\pi,0)$-$(0,\pi)$ are identical for pattern (a). 
(Thus they are always overlapped.)
ii. The spectra along the path $(-\pi,0)$-$(0,\pi)$-$(\pi,0)$ are 
symmetric about point $(0,\pi)$ for pattern (b), while
it is not for pattern (a). This can be also seen from Fig. \ref{Fig:FS}.
iii. For pattern (b), the two bands display no
overlap for $t'=0.1$ (in contrast to the case for $t'=0.15$). 
Then the ground state energy is simply a summation
for the `$-$' band, which becomes $t'$-irrelevant. So the 
optimal parameter $\delta^*$ should take the same value as that for $t'=0$,
as discussed in the earlier work \cite{Yuan}.

At last, we point out that the DOS in each dimerized state for finite $t'$ 
exhibits the feature of asymmetric double peaks. This could provide 
characteristic information on some measurable properties like optical 
absorption in real materials when available. 
%Moreover, an extra dip is present for pattern (b), but not for (a).

\section{Peierls transitions at $T>0$}
\begin{figure}
\epsfxsize=8.5cm
\epsfysize=5.5cm
\centerline{\epsffile{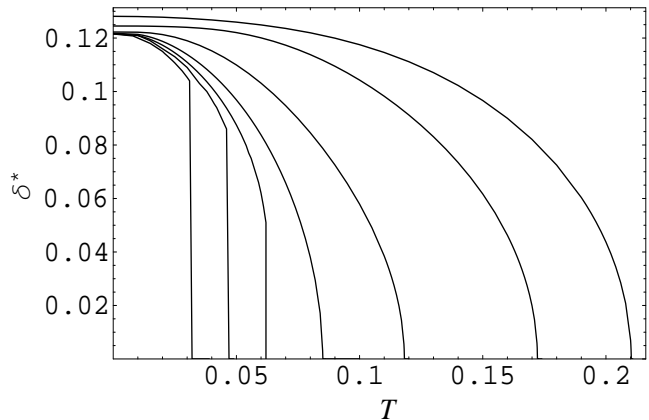}}
\medskip
\caption{The optimal dimerization parameter $\delta^*$ as a function of 
temperature $T$ under various values of $t'$ for pattern (a). 
The curves from right to left correspond to 
$t'=0,\ 0.12,\ 0.16,\ 0.168,\ 0.17,\ 0.171,\ 0.172$,
respectively.}
\label{Fig:Dels}
\end{figure}

\begin{figure}
\epsfxsize=8.5cm
\epsfysize=5.5cm
\centerline{\epsffile{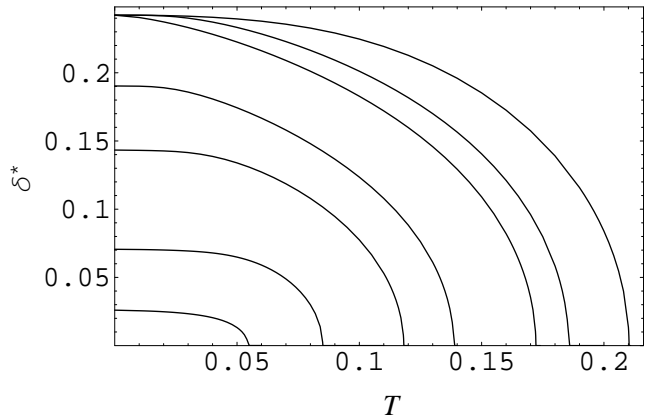}}
\medskip
\caption{The same as Fig. \ref{Fig:Dels} for pattern (b).
The curves from right to left correspond to 
$t'=0,\ 0.1,\ 0.12,\ 0.15,\ 0.16,\ 0.168,\ 0.17$,
respectively.}
\label{Fig:Dela}
\end{figure}

We discuss finite temperature transitions in this section. For clearness
we first look at the results for the two cases of pattern (a) and (b) 
separately, and then consider the competition between them. 
The optimal dimerization 
parameters with change of temperature were calculated for various 
values of $t'$. The results for pattern (a) are shown in Fig. \ref{Fig:Dels}
and those for pattern (b) in Fig. \ref{Fig:Dela}. 
A notable common property seen from both figures is that the Peierls 
transition temperature (denoted as $T_p^a$ and $T_p^b$ for pattern (a) and
(b), respectively) decreases with increase of $t'$. This just means that
the n.n.n. hopping $t'$ is unfavorable to the Peierls instability
to occur, which is expected and consistent with the results at ground state. 
It deserves to be pointed out that the lowering of the transition 
temperature $T_p$ with increasing $t'$ is consistent with 
the experimental findings in 
the bronzes (PO$_2$)$_4$(WO$_3$)$_{2m}$, where $T_p$ was found to
increase with $m$ because of better nesting properties \cite{Hess}.
Then we continue to look at the types of transitions in both
cases. For pattern (b), the optimal dimerization parameter $\delta^*$ 
goes to zero smoothly with increase of
$T$ for each $t'$, i.e., the Peierls transition is always of second-order.
However, the type of transition is variant for pattern (a). 
It is found that for most of $t'$ values the transion is of 
second-order, while when $t'$ is larger than about $0.17$ it becomes of 
first-order. The explanation for the first-order here is similar to
that for the first-order suppression by $t'$ at $T=0$, which
was presented in Ref. \cite{Yuan}.

\begin{figure}
\epsfxsize=8cm
\epsfysize=9.5cm
\centerline{\epsffile{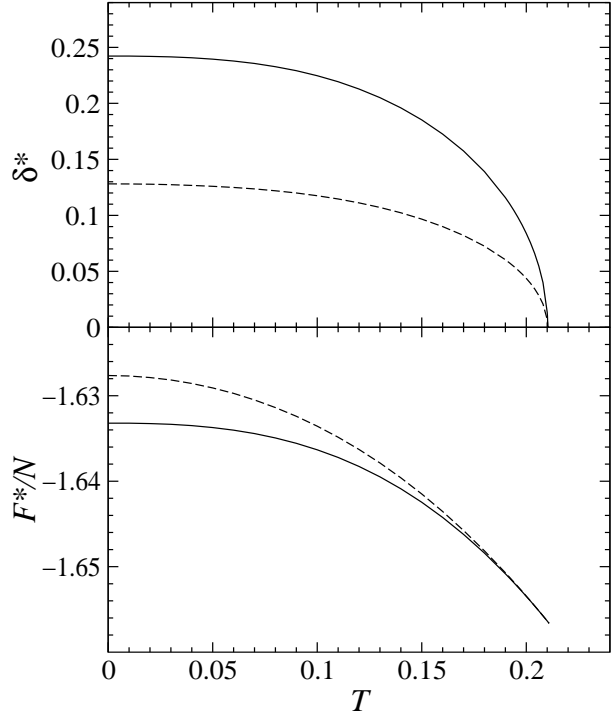}}
\medskip
\caption{The optimal values $\delta^*$ (upper panel) and the corresponding 
free energies $F^*$ (lower panel) as a function of $T$ for $t'=0$.
The dashed lines are for pattern (a) and the solid lines are for pattern (b).}
\label{Fig:Dtd0}
\end{figure}

\begin{figure}
\centerline{\epsfig{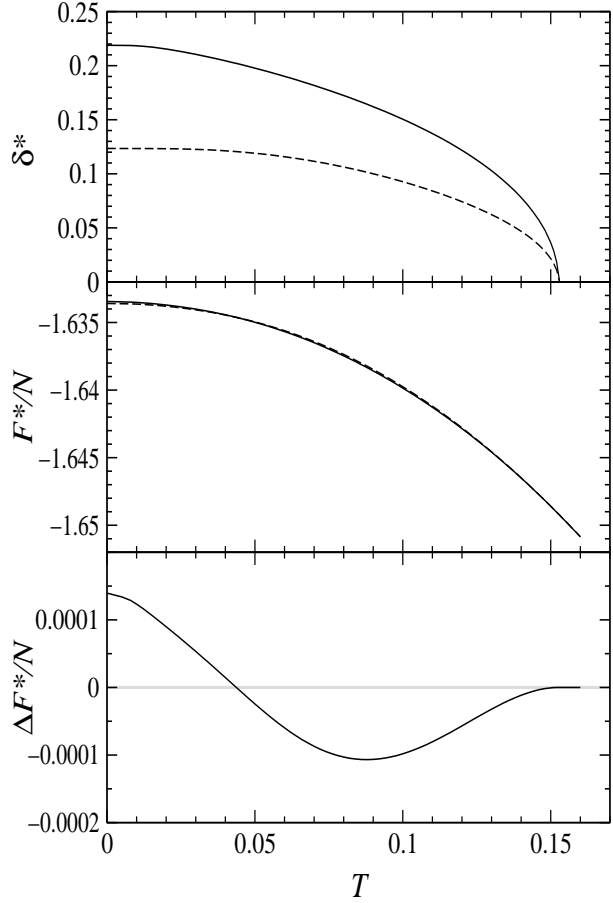}}
%\centerline{\epsffile{Dtd014.eps}}
\medskip
\caption{Similar to Fig. \ref{Fig:Dtd0} but with an additional plot
$\Delta F^*=F_b^*-F_a^*$ vs. $T$, for $t'=0.14$.}
\label{Fig:Dtd014}
\end{figure}

\begin{figure}
\centerline{\epsfig{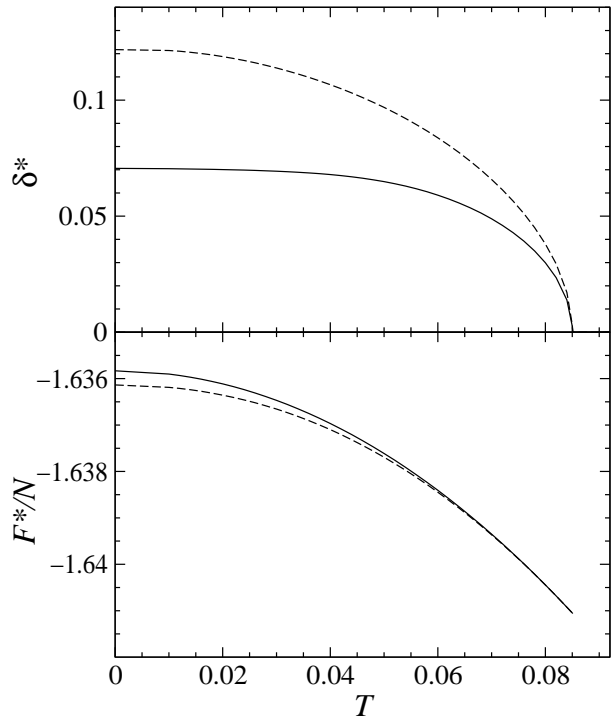}}
%\centerline{\epsffile{Dtd0168.eps}}
\medskip
\caption{The same as Fig.\ref{Fig:Dtd0} but for $t'=0.168$.}
\label{Fig:Dtd0168}
\end{figure}

\begin{figure}[tp]
\centerline{\epsfig{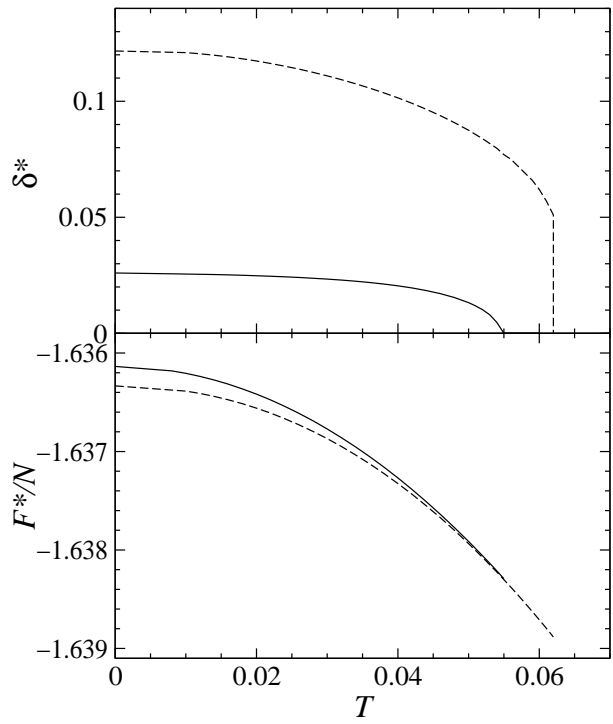}}
%\centerline{\epsffile{Dtd017.eps}}
\medskip
\caption{The same as Fig.\ref{Fig:Dtd0} but for $t'=0.17$.}
\label{Fig:Dtd017}
\end{figure}

To see the competition between the two dimerization patterns, 
one needs to compare the free energies $F^*$ for both patterns 
at their respective optimal dimerization values 
for each $t'$ and $T$. In the following we choose several $t'$ values 
$0,\ 0.14,\ 0.168,\ 0.17$ to show all different kinds of results, 
see Figs. \ref{Fig:Dtd0}-\ref{Fig:Dtd017}.  
In each figure the optimal dimerization
parameters and their corresponding free energies (per site) 
with change of temperature are compared for both patterns. 
We will discuss them in detail in the following.

For the cases of $t'=0,\ 0.14,\ 0.168$, 
it is always found that the transition temperatures for both patterns 
are identical, i.e., $T_p^a=T_p^b=T_p$, but the comparative results on 
free energies are different. For $t'=0$ the free energy for 
pattern (b) $F_b^*$ is always lower than
that for pattern (a) $F_a^*$ in the whole temperature region $T<T_p$
(see Fig. \ref{Fig:Dtd0}),
which means that the dimerized state with pattern (b) is always
prefered; while for $t'=0.168$ the result is exactly the reverse
(see Fig. \ref{Fig:Dtd0168}).
The most interesting case is that for the intermediate $t'=0.14$, 
see Fig. \ref{Fig:Dtd014}. To more clearly compare the magnitudes of 
$F_a^*$ and $F_b^*$, we have plotted their
difference $\Delta F^*=F_b^*-F_a^*$ as a function of temperature
in the lowest panel, which shows that $F_b^*$ is lower than 
$F_a^*$ in the intermediate 
temperature region and then becomes higher for low temperatures. 
Note that the absolute numerical error for the free energy in our calculations
is less than $10^{-8}$, so the minor difference $\Delta F^*$ is not 
an artifact. Therefore, in this case the transitions with decreasing 
temperature may be described as follows: 
at some critical temperature $T_c^1$ the system goes through 
a Peierls instability into the dimerized state (b). When temperature continues 
to decrease down to another critical value $T_c^2$ the system will 
turn from the dimerized state (b) into (a). That is to say, 
in the whole temperature region two successive Peierls transitions 
take place, as expected in the Introduction.

In addition, for the case of $t'=0.17$, the transition temperature 
$T_p^a$ becomes different from $T_p^b$ and the former is larger, see 
Fig. \ref{Fig:Dtd017}. In addition, the
free energy $F_a^*$ is always lower than $F_b^*$. Thus the pattern (a) 
is always dominant once $T<T_p^a$.\\

\newpage

\begin{figure}[tp]
\epsfxsize=8.5cm
\epsfysize=5.5cm
\centerline{\epsffile{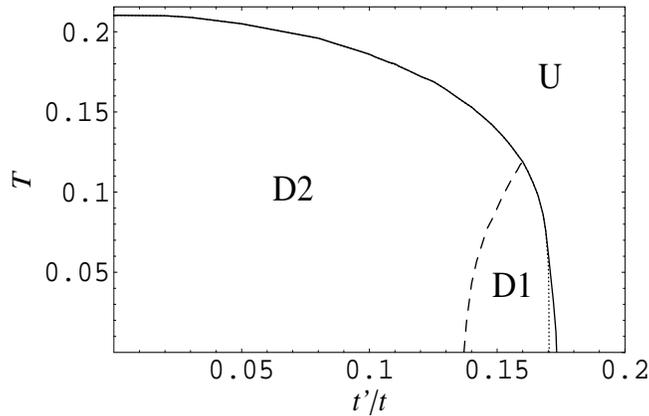}}
\medskip
\caption{The phase diagram for different stable states in the parameter space
of $t'$ and $T$. U,\ D1,\ D2 represent the uniform state, dimerized state with
pattern (a), dimerized state with pattern (b), respectively. The solid
and dotted lines are the phase boundaries between U and D1, and D2, 
respectively, when the two dimerization patterns are considered separately. 
Note that they are identical for most values of $t'$. 
The dashed line differentiates the stable state between D1 and D2.
In the narrow area between the solid and dotted line only the D1 state survives.}
\label{Fig:Phase}
\end{figure}

With the known results for several typical $t'$ values, we now construct the
phase diagram for different stable states in the full parameter space 
of $t'$ and $T$. The result is presented in Fig. \ref{Fig:Phase}, 
in which most of the above results have been summarized. 
From Fig. \ref{Fig:Phase} the transition temperatures 
$T_p^a$ (solid line) and $T_p^b$ (dotted line) with change of $t'$, 
as well as the competition between two 
dimerization patterns are clearly seen. Note that the solid
and dotted lines are identical for most values of $t'$, and become
separate only for a narrow region with large $t'$. 
%Also note that the abscissa and ordinate have been elaborated by the 
%previous Fig. \ref{Fig:DT0} and Fig. \ref{Fig:Dtd0}, respectively.
The most exciting result seen from Fig. \ref{Fig:Phase} is that the 
double Peierls transitions, i.e., the system
passing the states U$\rightarrow$D2$\rightarrow$D1 with decrease of
temperature, may take place for $t'$ in the region with approximately
$0.137<t'<0.16$.

\section{Discussion and Conclusion}

In the previous sections, extensive results about the Peierls instability,
including the rich phase diagram in the parameter space of $t'$ and $T$
have been discovered. Now we discuss the possible implications of
these results to real materials. Although our model may be
too simplified when fitting to real quasi-2D Peierls materials,
the results obtained are still instructive for the understanding of some 
experimental findings. (Actually some of instructions have already been 
presented in the suitable context above.)
For real materials, the n.n.n. hopping $t'$ is often indispensable,
and moreover, it can be changed by some external conditions, e.g., pressure. 
Depending on whether the ratio $t'/t$ is enhanced or reduced by the pressure,
the Peierls transition temperature is expected to decrease (and even vanish)
or increase by it. Experimentally both of possibilities were observed in 
the real quasi-2D bronzes, see Ref. \cite{Beille} and references therein.
Our results also suggest that applying pressure (if leading to an enhancement
of $t'/t$) may prevent the materials from the Peierls instability, and then 
possibly help them enter into the competing superconducting state. 
Interestingly for the bronze Li$_{0.9}$Mo$_6$O$_{17}$, it was found that the 
superconducting transition temperature is largely increased by pressure, 
associated with a sharp
decrease of the transition temperature of a possible Peierls 
instability \cite{Escribe}. In particular, our results show two successive
Peierls transitions with temperature, a phenomenon which was found in
TlMo$_6$O$_{17}$ \cite{Qin} and (PO$_2$)$_4$(WO$_3$)$_{2m}$ 
\cite{Schlenker,Wang,Hess,Beille}.
Essentially, the theoretical and experimental results are similar,
i.e., they both originate from the competition between two different lattice
distortion patterns. Of course, the real situations are more complicated
mainly because of intricate lattice distortions. Nevertheless, 
our work provides a good example in two dimensions to 
show double Peierls transitions at finite temperatures. Note that,
within the intrinsic limitations (i.e., adiabatic phonons and limited lattice
distortion patterns), all results presented here are exact 
in the thermodynamic limit.

Coming back to the region of Fig. \ref{Fig:Phase} where successive 
transitions U$\rightarrow$D2$\rightarrow$D1 take place, 
we further discuss an interesting issue stated as follows. In this
figure, it is obvious that the second transition D2$\rightarrow$D1 
is of first-order. Are there some intermediate states which may smoothly connect
D2 and D1? Actually, as a more general dimerization pattern, 
the lattice distortions may be assumed in the way:
$$u_{i,j}^x-u_{i+1,j}^x=(-1)^{i+j}u_x,\ \
  u_{i,j}^y-u_{i,j+1}^y=(-1)^{i+j}u_y
$$
with a ratio $r=u_y/u_x \in [0,1]$. The patterns (a) and (b)
correspond to the two limiting cases: $r=1$ and $0$, respectively.
If this generalization is included in the calculation, preliminary results 
give the following scenario with decreasing temperature: 
first, the transition still
happens between the uniform state and the dimerized state (b) 
(i.e., $r=0$). This state persists for a range of $T$ and then 
begins to turn continuously into other dimerized states with $r>0$. 
Depending on the value of $t'$, the $r=1$ state, i.e., the dimerized state (a)
may be quickly reached or may not be reached until $T=0$.
Now, the second transition between D1 and other dimerized states 
becomes of second-order.
Detailed results need a more careful calculation.

Finally, we briefly discuss the effects of electron correlations, 
e.g., the Hubbard on-site $U$ at ground state.
For $t'=0$, it was found
that the on-site Coulomb interaction is unfavorable to 
dimerization in two dimensions as soon as $U$ is present 
\cite{Tang,Mazumdar,Kopp}.
This is because the on-site $U$ favors the appearance of antiferromagnetic 
(AF) order for the 2D half-filled model, while the dimerization 
tends to stabilize local spin singlets. Thus both, Peierls and AF 
instabilities, will compete with each other in the presence of 
finite $\eta$ and $U$ \cite{Kopp}. On the other hand, the hopping $t'$, 
which breaks the perfect nesting, will frustrate the Peierls instability 
as shown here and also the AF instability as shown 
by several authors \cite{Hai}, when each of them is discussed individually.
Then it surely has nontrivial effects on the competition between
these two instabilities when they are considered jointly.
As far as the Peierls instability is concerned, it is argued that
both $t'$ and $U$ may cancel in part their effects 
when acting simultaneously, although each of them separately 
tends to suppress it. This topic will be left for future investigation.

In conclusion, the Peierls instabilities for a half-filled 
2D $t$-$t'$ model are thoroughly studied with consideration of 
two dimerization patterns (a) and (b). The effect of imperfect nesting
introduced by $t'$ on the Peierls instability, the properties of the
dimerized ground state, as well as the competition between two 
dimerized states are completely discussed. We have found that 
the n.n.n. hopping $t'$ will frustrate the Peierls instability 
for each of the dimerized states when considered separately.
Moreover, when the two dimerized states are considered
together, they will compete to result in a rich phase diagram in the 
parameter space of $t'$ and $T$. Prominently, double Peierls transitions, 
that is, the system passing
from the uniform state to one dimerized state and then to the other 
may take place with decrease of temperature. 

\section*{Acknowledgements}

The author would like to thank T. Kopp, T. Nunner, J. Shi, and S. Q. Shen
for valuable discussions.
This work was financially supported by the Deutsche Forschungsgemeinschaft
through SFB 484, the BMBF 13N6918/1, and the National Natural Science 
Foundation of China.

%end{references}
\end{document}